\documentclass[pra,twocolumn,preprintnumbers,superscriptaddress]{revtex4}
\usepackage{graphicx}
\usepackage{dcolumn}
\usepackage{bm}
\usepackage{graphicx}  
\usepackage{dcolumn}   
\usepackage{bm}        
\usepackage{verbatim}   
\usepackage[usenames]{color}  
\usepackage{stmaryrd}
\usepackage{float}
\usepackage{units}
\usepackage{textcomp}
\usepackage{textcomp}
\usepackage{amsmath}
\usepackage{commath}
\usepackage{amssymb}
\usepackage{esint}
\usepackage{ae,aecompl}
\usepackage[colorlinks=true,linkcolor=blue,citecolor=blue]{hyperref}

\begin{document}

\title{Interminiband absorption in a quantum ring superlattice in magnetic field with periodic vector potential}

\author{Vigen Aziz-Aghchegala}
\affiliation{Department of Physics, Urmia University of Technology, Urmia, Iran}

\author{Vram Mughnetsyan}
\email{vram@ysu.am}
\affiliation{Department of Solid State Physics, Yerevan State University, Alex Manoogian 1, 0025 Yerevan, Armenia}

\author{Ara Atayan}
\affiliation{Department of Medical Physics, Yerevan State Medical University, Koryun str. 2, 0025 Yerevan, Armenia}

\author{Albert Kirakosyan}
\affiliation{Department of Solid State Physics, Yerevan State University, Alex Manoogian 1, 0025 Yerevan, Armenia}

\begin{abstract}
The miniband Aharonov-Bohm oscillations and the interminiband absorption coefficient have been considered theoretically for one-layer superlattices of square and rectangular symmetries, composed of cylindrical quantum rings in the external transverse magnetic field with a periodic vector potential by the lattice constants. The crossings and anticrossings of the energies corresponding to different values of quasimomentum are observed. It is shown that the energy gap between the minibands and the sequence of the energies in each miniband can be tuned by the magnetic field. The interminiband absorption coefficient qualitatively depends on the symmetry of the superlattice, magnetic field induction and the incident light polarization. The obtained results indicate on the possibility to control the electronic and optical characteristics of the devices based on quantum ring superlattices.


\end{abstract}



\maketitle

\section{Introduction}
Nanostructured materials in which the system of conduction electrons can be considered as two-dimensional are subject of an extensive investigation during the last two decades. Among them graphene \cite{AokiBook,Geim,Novoselov} and other atomic monolayers are of great interest.
On the other hand, one of the promising courses of development of nowaday optoelectronics is the transition to zero dimensional nanostructures, such as quantum dots (QDs) \cite{ChakrabortyBook} and quantum rings (QRs) \cite{FominBook}. In contrast to QDs, QRs allow to observe experimentally the effects based on the quantum phase coherence such as Aharonov-Bohm effect. Intermediate band solar elements, heterojunction lasers and other optoelectronic devices based on QD and QR systems possess a number of advantages: temperature stability, wide spectral range, small dark current, high signal-to-noise ratio, etc. \cite{Ledentsov,Dai}. That is why the realization of ordered structures, such as two-dimensional superlattices (SL), composed of QDs \cite{Nakamura,Kohmoto} and QRs with almost the same size and shape is an important technological task \cite{Wu,Huang1}. This fact stimulated wide range of investigations on the effect of external fields on optoelectronic properties of quasi-two-dimensional electrons modulated by a periodic electrostatic potential \cite{Silberbauer,Gudmundson1,Gudmundson2,Barseghyan,Chakraborty1,Harutyunyan}.

The intraband optical absorption in QRs has attracted an enormous interest in recent years \cite{Baskoutas,Huang2,Xie} because of large optical nonlinearity observed in these structures and potential applications in photodetectors and high-speed electro-optical devices \cite{Caposso,Miller,Linares-Garcia,Chakraborty2}.

Since the appearance of the works of Azbel \cite{Azbel} and Hofstadter \cite{Hofstadter}, the study of electronic spectra in a two-dimensional lattice under the influence of a transverse (perpendicular to the lattice plane) magnetic field has become a field of continuous interest \cite{Gudmundson2,Ye,Gumbs,Zhang1,Guil,Ibrahim,Zhang2,Janecek,Dios-Leyva}. Recently the consideration of two-dimensional electrons in a spatially periodic transverse magnetic field or periodic magnetic potential is of special interest \cite{Gumbs,Zhang2,Xinyu,Lu,Ramezani}. In Ref. \cite{Gumbs} the static magneto-conductivity components are calculated depending on periodically modulated magnetic field for a square lattice. In Ref. \cite{Zhang2} the electronic band structure of GaAs/AlGaAs superlattice in a periodic magnetic field has been investigated.
A general scheme for synthesizing a spatially periodic magnetic field, or a magnetic lattice is presented in Ref. \cite{Xinyu}. A one dimensional SL on graphene with periodic Kronig-Penney model of magnetic potential which corresponds to periodic $\delta$-function-like magnetic field has been considered in Ref. \cite{Ramezani}. 
The above mentioned works indicate on the importance to study the effect of periodic magnetic field on 2D electrons and on the possibility to design devices with transport and optical properties which can be tuned by means of spatially modulated magnetic field.

In the present work we have considered the miniband Aharonov-Bohm oscillations and interminiband light absorption in InAs/GaAs quasi-two-dimensional QR SLs with square and rectangular symmetries subjected to a quasi-homogeneous transverse magnetic field with spatially periodic vector potential.

\section{Theory}
\begin{figure}
\centerline{\includegraphics[width=6cm]{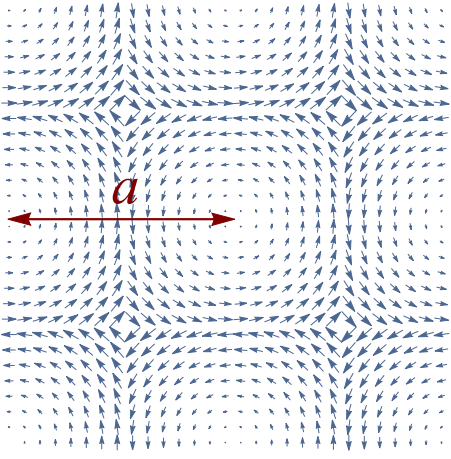}} \caption{(Colour on-line) Schematic view of the vector potential distribution in a SL of square symmetry with lattice constant $a$.}
\end{figure}
One-electron Hamiltonian in a QR SL in the presence of a transverse magnetic field with vector potential $\vec{A}$ can be written in the following form:
\begin{eqnarray}
\label{Hamiltonian}
\mathcal{H}=\frac{1}{2}\left(\hat{p}-\frac{e}{c}\vec{A}\right)\frac{1}{m(\vec{r})}\left(\hat{p}-\frac{e}{c}\vec{A}\right) \\ \nonumber
+\frac{1}{2}\texttt{g}(\vec{r})\mu_{B}\sigma_{z}(\nabla \times \vec A)_{z}+V(\vec{r}),
\end{eqnarray}
where $\sigma_z$ is the Pauli ``$z$" matrix,
$V(\vec{r})$ is the periodic potential of QR SL, which is taken to be $0$ inside the QRs and $V_{0}$ outside them, $m(\vec{r})$ and $\texttt{g}(\vec{r})$ are the electron effective mass and Land\'{e} $\texttt{g}$-factor, respectively, which by enough high accuracy can be taken to be equal to their values in the InAs inside the QRs and to their values in the GaAs outside the QRs, $\mu_{B}$ is the Bohr magneton. We assume that the width of the SL $d$ is small enough, leading to a strong quantization in the direction perpendicular to the SL's plane. This fact allows one to consider the in-plane motion of the electron independent on transverse one.

The Cartesian components of the vector potential are taken in the following periodic forms:
\begin{equation}
\label{VecPot}
\begin{cases}
A_{x}=-\dfrac{B}{2}\displaystyle\sum\limits_{C_{y}}(y-C_{y})\theta\left(\dfrac{a_{y}}{2}-|y-C_{y}|\right), \\
A_{y}=\dfrac{B}{2}\displaystyle\sum\limits_{C_{x}}(x-C_{x})\theta \left(\dfrac{a_{x}}{2}-|x-C_{x}|\right), \\
A_{z}=0,
\end{cases}
\end{equation}
where $a_{x(y)}$ is the lattice constant in $x(y)$ direction, $B$ is the magnitude of magnetic field induction and the summations are carried out by the coordinates of the QRs centers $C_{x}$ and $C_{y}$. As is known the Hamiltonian with the vector potential of usual symmetric gauge ($A_{x}=-By/2$, $A_{y}=Bx/2$) does not commute with translation operators of the SL resulting to a fractal structure for the energy spectrum \cite{Hofstadter,Ferrari}. In contrast, the Hamiltonian (\ref{Hamiltonian}) with vector potential (\ref{VecPot}) commutes with corresponding translations allowing the existence of continuous minibands in $k$-space, which are easier to control by external factors. The schematic view of the distribution of vector potential given by Eq. (\ref{VecPot}) in a SL of square symmetry is presented in Fig.1. It is noteworthy, that the magnetic flux of the considered field is 0 in each of the unite cells if one includes the edges of the unite sell in the calculation of flux. When the edges are not included the magnetic flux is $BS_{0}$ ($S_{0}$ is the area of the unite cell) as for the homogeneous magnetic field (see the Appendix A).

Due to the periodicity of the Hamiltonian (\ref{Hamiltonian}) one can make
the following Fourier transformations:
\begin{eqnarray}
\label{FPsi}
\psi(\vec{r})=S^{-1}e^{i\vec{k}\vec{r}}U_{\vec{k}}(\vec{r})=\frac{1}{S}\sum_{\vec{g}}u_{\vec{k},\vec{g}} e^{i (\vec{k}+\vec{g})\vec{r}},
\\
\label{FVm}
V(\vec{r})=\sum_{\vec{g}}V_{\vec{g}} e^{i \vec{g} \vec{r}},
\quad
\frac{1}{m(\vec{r})}=\sum_{\vec{g}}m^{-1}_{\vec{g}} e^{i\vec{g}\vec{r}},
\\
\label{FgA}
\texttt{g}(\vec{r})=\sum_{\vec{g}}\texttt{g}_{\vec{g}} e^{i \vec{g}\vec{r}},
\quad
\vec{A}(\vec{r})=\sum_{\vec{g}}\vec{A}_{\vec{g}} e^{i \vec{g} \vec{r}},
\\
\label{Ff1}
\frac{\vec{A}}{m(\vec{r})}\equiv{\vec{F}_{1}(\vec{r})}=\sum_{\vec{g}}\vec{F}_{1,\vec{g}}e^{i \vec{g}\vec{r}},
\\
\label{Ff2}
\frac{A^{2}}{m(\vec{r})}\equiv{F_{2}(\vec{r})}=\sum_{\vec{g}}F_{2,\vec{g}}e^{i \vec{g}\vec{r}},
\end{eqnarray}
where $U_{\vec{k}}(\vec{r})$ is the Bloch amplitude, $\vec{k}$ is the quasimomentum, $S$ is the effective area of the SL. The Fourier transforms in Eqs. (\ref{FPsi})-(\ref{Ff2}) can be expressed as follows:
\begin{widetext}
\begin{eqnarray}
\label{Vg}
V_{\vec{g}}=V_{0}\Bigg(\delta_{\vec{g},0}-\frac{2\pi(R_{2}J_{1}(R_{2}g)-R_{1}J_{1}(R_{1}g))}{S_{0}g}\Bigg),
\\
\label{mg}
m^{-1}_{\vec{g}}=\frac{\delta_{\vec{g},0}}{m_{GaAs}}+
\frac{2\pi(R_{2}J_{1}(R_{2}g)-R_{1}J_{1}(R_{1}g))}{S_{0}g}
\left(\frac{1}{m_{InAs}}-\frac{1}{m_{GaAs}}\right),
\\
\label{Landeg}
\texttt{g}_{\vec{g}}=\texttt{g}_{GaAs}\delta_{\vec{g},0}+
\frac{2\pi(R_{2}J_{1}(R_{2}g)-R_{1}J_{1}(R_{1}g))}{S_{0}g}
\left(\texttt{g}_{InAs}-\texttt{g}_{GaAs}\right),
\\
\label{f1g}
F_{1,\vec{g},x(y)}=+(-)\frac{i B}{2}\Bigg[\frac{2\pi g_{y(x)}(R_{2}^{2}J_{2}(R_{2}g)-R_{1}^{2}J_{2}(R_{1}g))}
{S_{0}g^{2}}\left(\frac{1}{m_{InAs}}-\frac{1}{m_{GaAs}}\right) \nonumber \\
-\frac{\delta_{g_{x(y)},0}((-1)^{n_{y(x)}}-\delta_{g_{y(x)},0})}{m_{GaAs}g_{y(x)}}\Bigg],
\end{eqnarray}

\begin{eqnarray}
\label{f2g}
F_{2,\vec{g}}=\frac{B^{2}}{2}\Bigg[\pi \left(2\frac{R_{2}^{2}J_{2}(gR_{2})-R_{1}^{2}J_{2}(gR_{1})}{S_{0}g^{2}}-\frac{R_{2}^{3}J_{3}(gR_{2})-R_{1}^{3}J_{3}(gR_{1})}{S_{0}g}\right) \nonumber \\ \times\left(\frac{1}{m_{InAs}}-\frac{1}{m_{GaAs}}\right)+
\frac{1}{m_{GaAs}}\left(\frac{(-1)^{n_{y}}\delta_{g_{x},0}}{g_{y}^{2}}+\frac{(-1)^{n_{x}}\delta_{g_{y},0}}{g_{x}^{2}}+
\frac{\delta_{\vec{g},0}(a_{x}^{2}+a_{y}^{2})(\pi^{2}-2)}{8\pi^{2}}\right)
\Bigg],
\\
\label{Ag}
A_{g_{x(y)}}=-(+)\frac{i B \delta_{g_{x},0}}{2}\frac{(-1)^{n_{y(x)}}-\delta_{n_{y(x)}}}{g_{y(x)}}.
\end{eqnarray}
\end{widetext}

In Eqs. (\ref{Vg})-(\ref{Ag}) $J_{i}(\xi)$ is the first kind Bessel function of the $i$-th order, $S_{0}$ is the area of the SL's unite cell, $R_{1}$ and $R_{2}$ are the inner and the outer radii of the QRs, respectively \cite{Mughnetsyan1}, $\delta_{\xi,\zeta}$ is the Kronecker delta, $g_{x(y)}=2\pi n_{x(y)}/a_{x(y)}$ is the Cartesian component of the reciprocal lattice vector $\vec{g}$, $(n_{x(y)}=0,\pm1,\pm2,...)$, $g^{2}=g^{2}_{x}+g^{2}_{y}$. For the convenience the Fourier transformation for the functions $\vec{F}_{1}(\vec{r})$ and $F_{2}(\vec{r})$ which figure in Hamiltonian after the simplification of (\ref{Hamiltonian}) is also made as it is presented in Eqs. (\ref{Ff1}) and (\ref{Ff2}).
Substituting the expressions (\ref{FPsi})--(\ref{Ff2}) to the Ben
Daniel-Duke's equation $\mathcal{H}\psi=E\psi$ for each spin polarized state (there is no mixing between the spin states) one can arrive to the following set of linear equations:
\begin{widetext}
\begin{eqnarray}
\label{SetEq}
\begin{aligned}
\sum_{\vec{g}^{\prime}}\Bigg(\frac{\hbar^{2}}{2}m^{-1}_{\vec{g}-\vec{g}^{\prime}}(\vec{k}+\vec{g})(\vec{k}+\vec{g}^{\prime})
+V_{\vec{g}-\vec{g}^{\prime}}-E\delta_{\vec{g},\vec{g}^{\prime}}-
\frac{\hbar e}{c}\vec{F}_{1,\vec{g}-\vec{g}^{\prime}}(\vec{k}+\vec{g}^{\prime})
+\frac{e^{2}}{2c^{2}}F_{2,\vec{g}-\vec{g^{\prime}}} +(-) E_{Z,\vec{g}}(B)\Bigg)u_{\uparrow(\downarrow)\vec{k},\vec{g}^{\prime}}=0,
\end{aligned}
\end{eqnarray}
\end{widetext}
were $E_{Z,\vec{g}}(B)=\texttt{g}_{\vec{g}-\vec{g}\prime}\mu_{B}\sigma_{z}B/2$ is the Fourier transform of the Zeeman therm and $u_{\uparrow(\downarrow)\vec{k},\vec{g}}$ is the Fourier transform of Bloch amplitude for the spin up (spin down) state. The energy values and the Fourier transforms of Bloch amplitude for
each fixed value of quasi-momentum $\vec{k}$ can be obtained by the
exact diagonalization of the set of equations (\ref{SetEq}).
\begin{figure}
\centerline{\includegraphics[width=8.9cm]{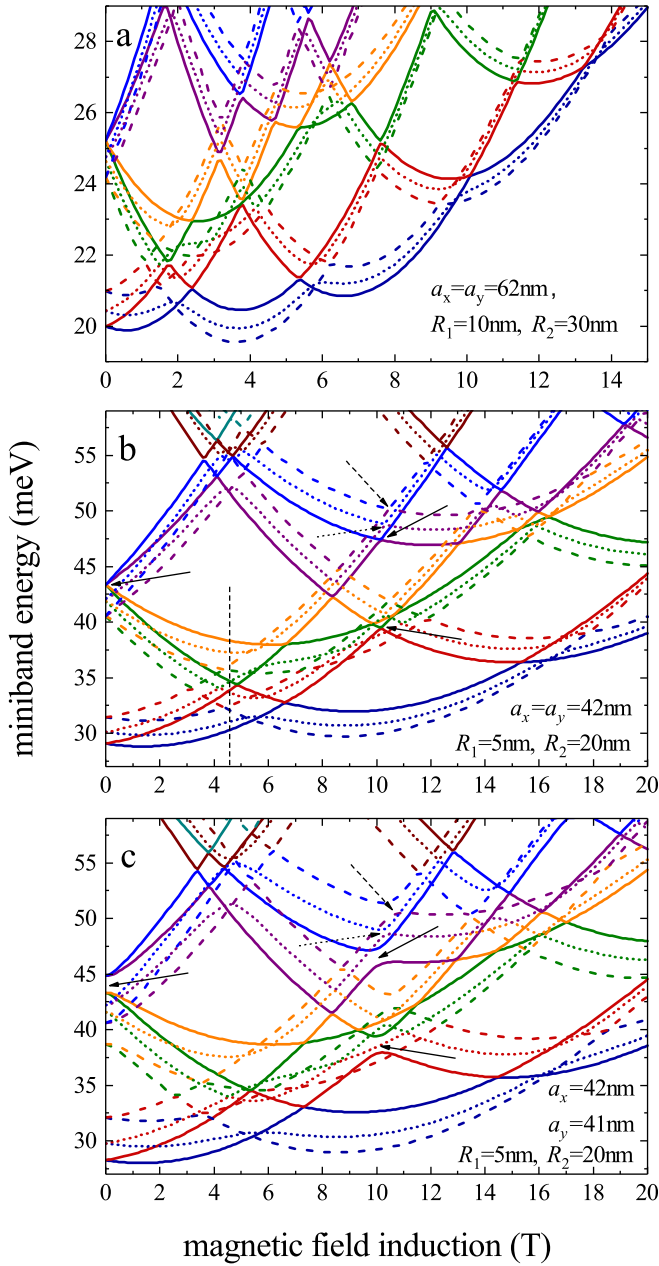}} \caption{(Colour on-line)
Dependence of electron energy on magnetic field induction for square (a,b) and rectangular (c) QR SL for three different values of the quasimomentum:
$k_{x}=k_{y}=0$ (solid lines), $k_{x}=\pi / a_{x}$, $k_{y}=\pi / a_{y}$ (dashed lines) and $k_{x}=\pi / 2a_{x}$, $k_{y}=\pi / 2a_{y}$ (dotted lines).}
\end{figure}

Taking into account that only the direct optical transitions between two minibands are allowed  \cite{Aziz-Aghchegala1}, the absorption coefficient (AC) in the dipole approximation can be expressed as follows:
\begin{eqnarray}
\label{AC}
\begin{aligned}
\alpha(\omega)=\alpha_{0} \delta_{s_{i},s_{f}}
\int \limits_{FBZ}d^{2}k|M_{i,f}(\vec{k})|^{2}\times \\
\delta(\hbar \omega-(E_{f}(\vec{k})-E_{i}(\vec{k}))
F(E_{i}(\vec{k}))(1-F(E_{f}(\vec{k})))&
\end{aligned}
\end{eqnarray}
where
\begin{equation}
\label{MatEl}
\begin{aligned}
M_{i,f}(\vec{k})=\hbar \sum \limits_{\vec{g}} u^{(i)}_{s_{i},\vec{k},\vec{g}}u^{(f)}_{s_{f},\vec{k},\vec{g}}(\vec{g}\vec{\eta})
\end{aligned}
\end{equation}
is the dipole matrix element of the transitions from the i-th to the f-th miniband, $u^{(l)}_{s_{l},\vec{k},\vec{g}}$ is the Fourier transform of the Bloch amplitude for the $l$-th miniband corresponding to the spin polarization $s_{l}$, $\delta_{s_{i},s_{f}}$ is the Kronecker delta for the initial and the final spin polarizations $s_{i}$ and $s_{f}$, $\alpha_{0}=e^{2}(m_{0}^{2}c d \omega \sqrt{\epsilon})^{-1}$,
$\omega$ and $\vec{\eta}$ are the frequency and the vector of polarization of incident photon, $F(E)$ is the Fermi-Dirac distribution function with chemical potential $\mu$ and absolute temperature $T$, $m_{0}$ and $e$ are the free electron mass and the charge, respectively, $\varepsilon$ is the dielectric constant of the media, $c$ is speed of light and the integration is carried out over the first Brillouin zone (FBZ).

\section{Results and discussion}

\begin{figure*}
\centerline{\includegraphics[width=10cm]{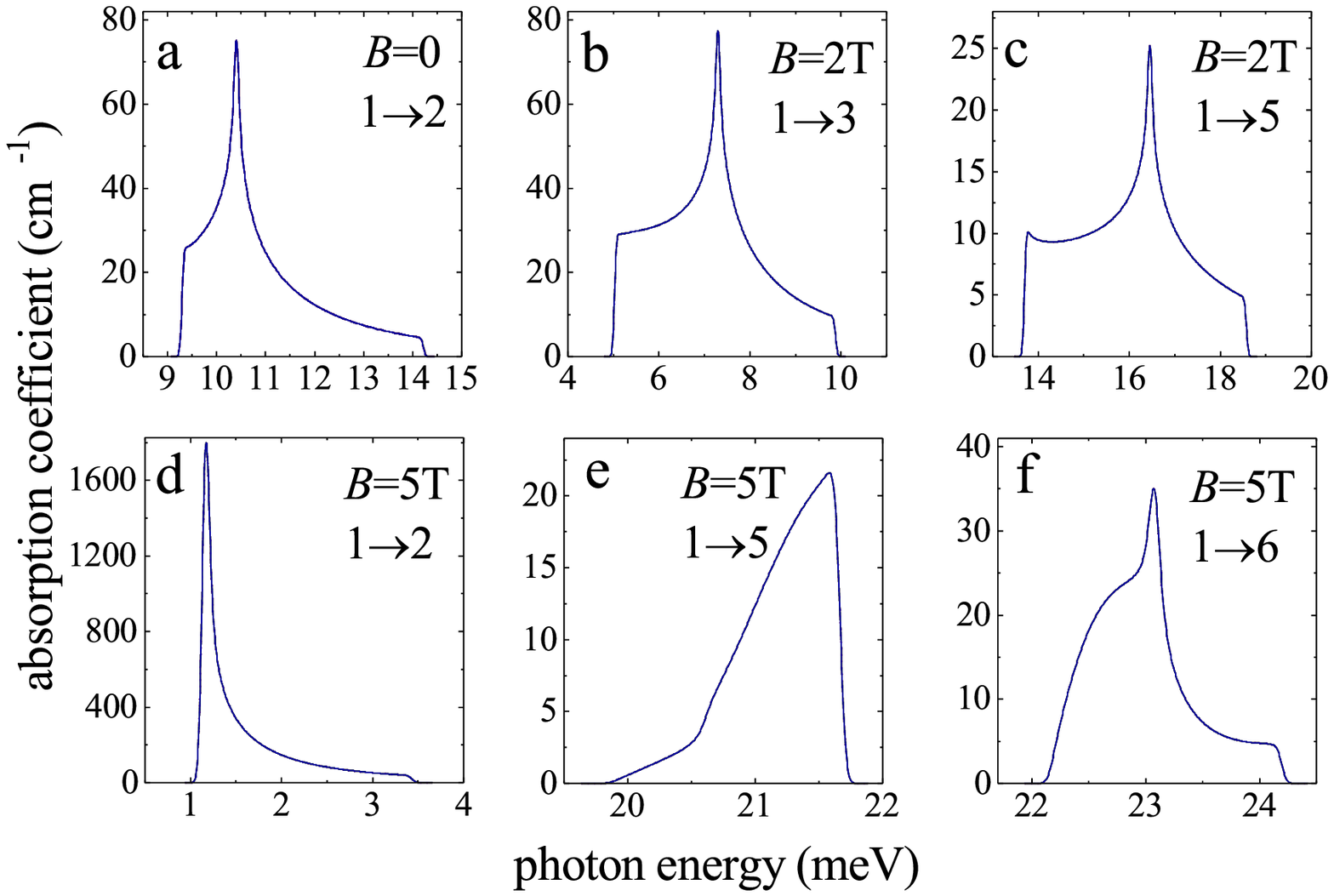}} \caption{(Colour on-line)
Dependence of AC on the incident photon energy at temperature near 0K for QR SL with lattice constants of $a_{x}=a_{y}=42$nm. The considered values of the light polarization angle are $\varphi=0$, $\pi/6$, $\pi/4$, $\pi/3$ and $\pi/2$.}
\end{figure*}
\begin{figure*}
\centerline{\includegraphics[width=19cm]{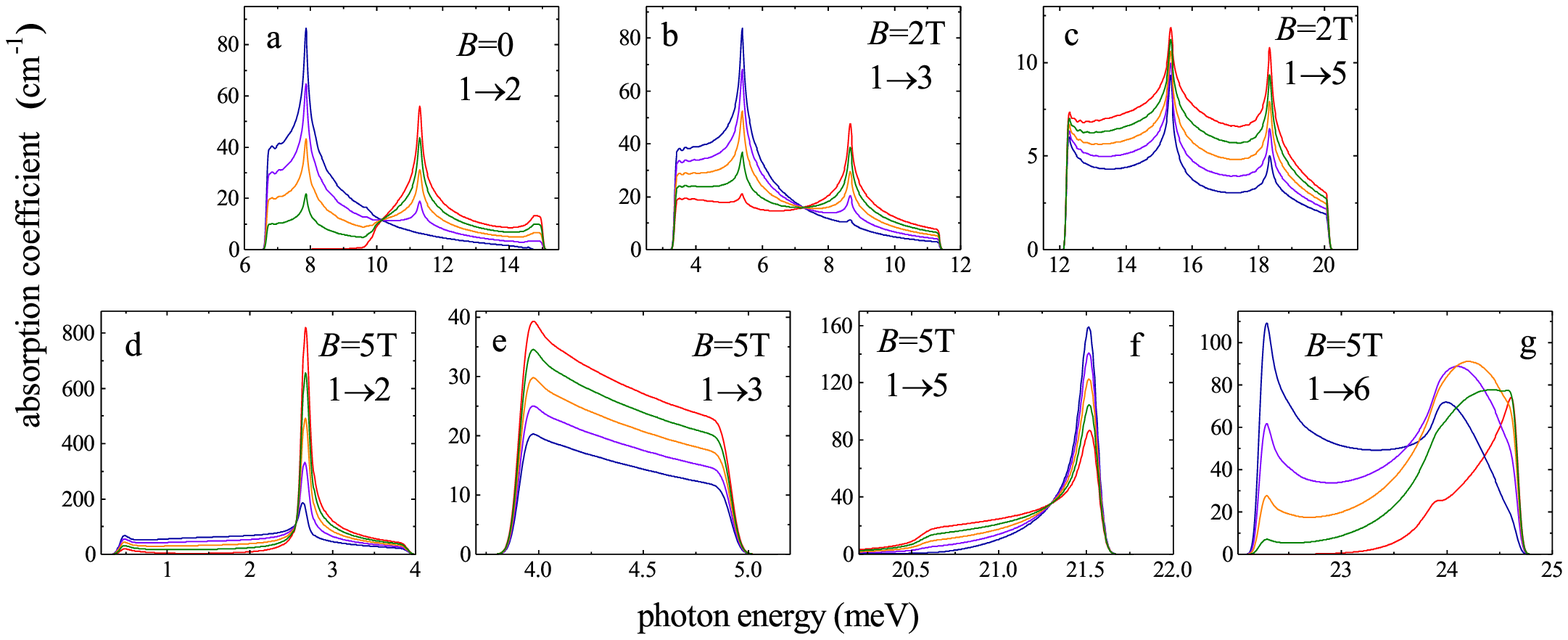}} \caption{(Colour on-line)
Dependence of the absorption coefficient on the incident photon energy at temperature near 0K for QR SL with lattice constants of $a_{x}=42$nm, $a_{y}=41$nm.  The considered values of the light polarization angle are $\varphi=0$ (red lines), $\pi/6$ (green lines), $\pi/4$ (orange lines), $\pi/3$ (violet lines) and $\pi/2$ (blue lines).}
\end{figure*}
\begin{figure}
\centerline{\includegraphics[width=7cm]{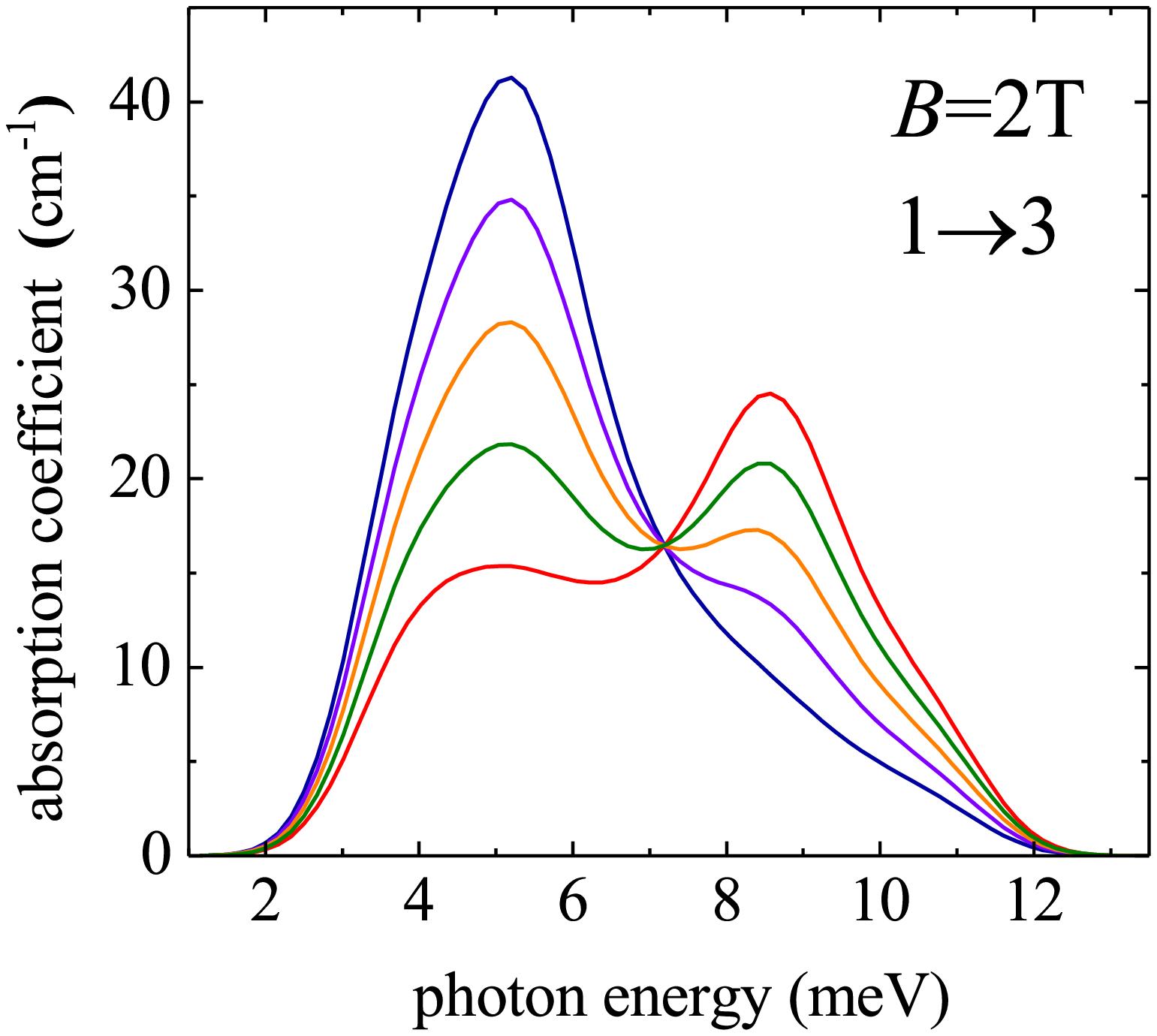}} \caption{(Colour on-line)
Dependence of the absorption coefficient on the incident photon energy at temperature $T=10$K for QR SL with lattice constants of $a_{x}=42$nm, $a_{y}=41$nm. The considered values of the light polarization angle are $\varphi=0$ (red lines), $\pi/6$ (green lines), $\pi/4$ (orange lines), $\pi/3$ (violet lines) and $\pi/2$ (blue lines).}
\end{figure}
Our calculations are carried out for InAs/GaAs QR SL. The value of the confining potential is taken to be $0$ inside the rings and $341$meV outside them \cite{Ghosh}. We use the values for the electron effective mass inside the QRs and outside them $m_{InAs}=0.026m_{0}$ and  $m_{GaAs}=0.067m_{0}$, respectively. Taking into account that for the considered values of parameters electron is mostly localized in the QR regions, for calculation of the absorption coefficient (\ref{AC}) we use the value for the dielectric constant of InAs material  $\epsilon=12.9$ \cite{Adachi}.

In Fig.2 the dependencies of the miniband energies on magnetic field induction at fixed points in the FBZ $k_{x}=k_{y}=0$, $k_{x}=\pi/a_{x},k_{y}=\pi/a_{y}$ and $k_{x}=\pi/2 a_{x},k_{y}=\pi/2 a_{y}$, are presented for SL of square symmetry (Fig.2a and b) and for SL of rectangular symmetry (Fig.2c). Two cases of SL of square symmetry, namely for the values of parameters  $R_{1}=10$nm, $R_{2}=30$nm, $a_{x}=a_{y}=62$nm (Fig.2 a) and $R_{1}=5$nm, $R_{2}=20$nm, $a_{x}=a_{y}=42$nm (Fig.2 b) have been considered. One can observe an oscillatory behaviour of the energy at each point of the FBZ due to Aharonov-Bohm effect in each QR composing the SL. Strictly speaking, there is no a certain value for the frequency of oscillations, because of the increase of the degree of electron's localization with increasing of magnetic field induction. However, from the comparison of Figs.2a and b it is clear that more frequent oscillations are observed for a SL composed of QRs with larger size (Fig.2a), because of smaller value of $B$ for which magnetic flux through each QR equals to the flux quantum $\phi_{0}=h c/e$. Crossings and anticrossings  for entire minibands (see for example the region where the solid, dashed and dotted lines are mentioned by the corresponding arrows in Fig.2b and c) as well as Zeeman splittings (for instance, the splitting of the lowest two minibands mentioned by red and blue lines for small values of $B$) have been observed. The change of the sequence of energies with regard to the values of quasimomentum due to miniband crossings is also obvious (Fig.2b). More detailed observation brings to light the following regularities. In the SL of square symmetry with lattice constants $a_{x}=a_{y}=a$ (Figs.2a and b) the levels corresponding to $\vec{k}=0$ (solid lines) or $k_{x}=k_{y}=\pi/a$ (dashed lines) behave similarly with ones for a single QR because of the vanishing group velocity in the center and the corners of the FBZ. Hence, it is possible to introduce a quantum number $L$ for each miniband which is analogue of the angular quantum number for quantum states in a single QR. One can assume that $L=0$ for the first miniband and $L$ gets positive (negative) integer values for the miniband energies which increase (decrease) when $B$ increases starting from $B=0$. However this description is not valid for the energy levels corresponding to $k_{x}=k_{y}=\pi/2a$ (dotted lines in Fig.2b). This levels behave similarly with ones in a single QR far from the regions of minibands crossing. When passing through these regions transition of dotted lines between the minibands corresponding to different values of $L$ takes place (for instance, in Fig.2b the yellow dotted line in the left side of the perpendicular dashed line belongs to the miniband corresponing to $L=-1$, while in the right side the same line belongs to the miniband which corresponds to $L=1$). It is noteworthy that some crossings of the energy levels for square SL (mentioned by arrows in Fig.2b) are transformed to anticrossings for rectangular SL (mentioned by arrows in Fig.2c). This transformation takes place for the levels with the same spin polarization. The minibands which show anticrossings correspond to the values of $L$ which differ by an even number (for example the anticrossings mentioned by solid arrows in Fig.2b correspond to the difference of $L$ by 2). So, the reduction of the SL symmetry from square to rectangular one leads to the splitting of the electronic minibands corresponding to the same parity of $L$ and the same spin polarization.

In Fig.3 the AC for QR SL with square symmetry is presented for different polarizations of the incident photon. It is assumed here (as well as in Fig.4) that the temperature is near $0$K, so the two lowest minibands with opposite spin polarizations are totally filled by electrons while the higher minibands are empty. It means that $F(E_{i})=1$ and $F(E_{f})=0$ in Eq. (\ref{AC}). A Gaussian distribution function ($(1/\sqrt{\pi}\Gamma)\exp(-(\hbar \omega- (E_{f}(\vec{k})-E_{i}(\vec{k})))^{2}/\Gamma^{2})$) has been used instead of the Dirac $\delta$-function in Eq. (\ref{AC}) with a very small value for the energy broadening parameter $\Gamma=0.05$meV \cite{Arzberger}.
The abrupt jumps of the curves correspond to the values of incident photon energy $\hbar \omega$ which are equal to the energy difference between the edges of minibands. Due to the symmetry of the energy dispersion surface in momentum space the AC is the same for all the considered values of the angle $\varphi$ between the light polarization vector and axis $x$ ($\varphi=0, \pi/6, \pi/4, \pi/3$ and $\pi/2$). When the magnetic field induction $B=0$ (Fig.3 a) the 1st and the 2nd minibands as well as the 3rd, 4th, 5th and the 6th minibands merge. That is why the only transitions between two minibands (referred as $1\rightarrow 2$ transitions) are considered. A single maximum is observed at the same value of the incident photon energy for all the considered directions of light polarization. For $B=2$T only $1\rightarrow 3$ and $1\rightarrow 5$ transitions are allowed, because of the same spin polarization of the states $1$, $3$ and $5$ (in the absence of the spin-orbit coupling spin is a ``good" quantum number). For $B=5$T the allowed transitions are $1\rightarrow 2$, $1\rightarrow 5$ and $1\rightarrow 6$. Comparison of Figs.3, a-f shows that the allowed transitions between the minibands as well as the magnitude and the frequency dependence of AC can be tuned efficiently by means of external magnetic field.

Fig.4 represents the dependencies of the AC on the incident photon energy for QR SL with lattice constants $a_{x}= 42$nm and $a_{y}= 41$nm for different values of magnetic field induction and different directions of the light polarization vector. As in the case of SL of square symmetry there are different transitions with non-vanishing matrix elements for different values of magnetic field induction. However, in contrast to SL of square symmetry, for some transitions there are two values of photon energy for which AC has obvious maxima (see Figs.4 a, b, d and f). The duplication of the absorption maximum is the result of the symmetry reduction of the dispersion surfaces. Note that there is a single maximum of AC for the angles of the polarization $\varphi=0$ (the right maximum) and $\varphi=\pi/2$ (the left maximum) even for the SL of rectangular symmetry. Interestingly, for some transitions there is a certain value of the incident photon energy when the AC does not depend on the light polarization (see Figs.4 a, b, d, and f) and the curves corresponding to different directions of polarization vector intersect. This value of photon energy corresponds to the section of the surface $E_{f}(\vec{k})-E_{i}(\vec{k})$ by the plane of constant energy $E=\hbar \omega$ which has a square symmetry in momentum space because only the values of $k$ which belong to the mentioned section contribute to AC. It is noteworthy the absence of the photon energy which corresponds to independence of AC on light polarization for $1\rightarrow5$ transitions when $B=2T$ (Fig.4c) and for $1\rightarrow3$, $1\rightarrow6$ transitions when $B=5T$ (Figs.4e and 4g). The reason is that in the mentioned cases not all the values of $k$ belonging to the section of $E_{f}(\vec{k})-E_{i}(\vec{k})$ by constant energy plane contribute to the AC.

Finally, Fig.5 represents the dependencies of the AC on the incident photon energy when the temperature of the system $T=10$K. Here the values of all other parameters coincide with ones in Fig.4 b. The value of the chemical potential in Fermi-Dirac distribution function $F(E)$ in Eq. (\ref{AC}) is obtained from the following assumptions: 1. the electrons' concentration does not depend on temperature due to the large gap between the valence and conduction bands, 2. the two lowest minibands with opposite spin polarizations are totally filled by electrons and all the higher minibands are empty at $T=0$K. As in Fig.4 b one can easily observe the intersection point of curves corresponding to different polarizations of incident photon. The comparison with Fig.4 b indicates on the reduction in the magnitude and the red-shift of the AC's maxima and on a considerable smearing of the curves due to the temperature effect.

\section{Conclusion}
In summary, we have considered the effect of the external magnetic field on the miniband Aharonov-Bohm oscillations and the interminiband light absorption coefficient for one-layer superlattices of square and rectangular symmetries composed of cylindrical quantum rings. The crossings and anticrossings as well as Zeeman splitting for single energy levels corresponding to certain values of quasimomentum as well as for entire minibands have been observed. It is shown that in SL of square symmetry the energy levels at the center or the corners of the first Brillouin zone behave similarly with ones for single quantum ring and can be described by a quantum number $L$ which is analogy of the angular quantum number. However the energies at intermediate points of the first Brilloun zone can be bundled with certain values of $L$ only in the regions far from the minibands' crossing. The reduction of the SL symmetry from square to rectangular one leads to the splitting of the electronic minibands corresponding to the same parity of $L$ and the same spin polarization.

The interminiband absorption coefficient has a single maximum and does not depend on the incident light polarization in a superlattice of square symmetry. However it significantly depends on magnetic field induction. The effect of the light polarization on the absorption coefficient is significant for superlattice of rectangular symmetry. For some transitions there are two values of photon energy where absorption coefficient has maximum and there is a value of photon energy were the curves corresponding to different polarizations of incident photon intersect. The reduction in the magnitude and the red-shift of the absorption coefficient's maxima and the considerable smearing of the absorption curves due to temperature is observed. The significant effect of the translational symmetry, magnetic field and the incident light polarization on the miniband energies and absorption coefficient makes possible the effective control of electro-optical characteristics of devises based on quantum ring superlattices.
\begin{acknowledgments}
This work was supported by the State Committee of Science of RA  (research project 18T-1C223). V.M. acknowledges partial financial support from EU H2020 RISE project CoExAN (Grant No. H2020-644076).
\end{acknowledgments}

\begin{widetext}
\appendix
\section{Derivation of the magnetic flux in the unite cell of superlattice}

Consider the flux of the magnetic field described by the vector potential (\ref{VecPot}) through the unite sell of the SL with lattice constants $a_{x}$ and $a_{y}$:
\begin{equation}
\label{fi}
    \varPhi=\int \limits_{-a_{x}/2}^{a_{x}/2}\int \limits_{-a_{y}/2}^{a_{y}/2}\bigg(\frac{\partial A_{y}}{\partial x}-\frac{\partial A_{x}}{\partial y} \bigg)dxdy=a_{y}\int \limits_{-a_{x}/2}^{a_{x}/2}\frac{\partial A_{y}}{\partial x}dx-
    a_{x}\int \limits_{-a_{y}/2}^{a_{y}/2}\frac{\partial A_{x}}{\partial y}dy.
\end{equation}
The first integral in the right hand side of Eq. (\ref{fi}):
\begin{equation}
\label{int}
    \int \limits_{-a_{x}/2}^{a_{x}/2}\frac{\partial A_{y}}{\partial x}dx=
    \frac{B}{2}\int \limits_{-a_{x}/2}^{a_{x}/2}\theta\bigg(\frac{a_{x}}{2}-|x|\bigg)dx-
    \frac{B}{2}\int \limits_{-a_{x}/2}^{a_{x}/2}|x|\delta\bigg(\frac{a_{x}}{2}-|x|\bigg)dx= \frac{B}{2}\bigg(a_{x}-
    2\int \limits_{0}^{a_{x}/2}x\delta\bigg(\frac{a_{x}}{2}-x\bigg)dx\bigg)=0.
\end{equation}
Analogically, one can show that the second term in the right hand side of Eq. (\ref{fi}) is also 0 and hence $\varPhi=0$. Note, that the contribution of the edges of unite cell is connected with integrals in Eq. (\ref{int}) containing Dirac delta-functions and the neglection of these integrals leads to the value of magnetic flux $\varPhi=BS_{0}$.
\end{widetext}

\end{document}